\documentclass[12pt]{article}
 \usepackage[cp1251]{inputenc} 
 \usepackage[english, russian]{babel} 
 \usepackage{amsmath, latexsym, amssymb, bm, array, graphics, eucal,
 amsfonts,}

 \renewcommand{\baselinestretch}{1.2}
\usepackage[dvips]{graphicx}

\begin{document}
\thispagestyle{empty}
\large
\renewcommand{\abstractname}{Abstract}
\renewcommand{\refname}{\begin{center}
 REFERENCES\end{center}}
\newcommand{\mc}[1]{\mathcal{#1}}
\newcommand{\E}{\mc{E}}
\makeatother

\begin{center}
\bf Boundary problems for the one-dimensional kinetic equation
with frequency of collisions, affine depending on the module
velocity
\end{center} \medskip

\begin{center}
  \bf
  A. L. Bugrimov\footnote{$fakul-fm@mgou.ru$},
  A. V. Latyshev\footnote{$avlatyshev@mail.ru$} and
  A. A. Yushkanov\footnote{$yushkanov@inbox.ru$}
\end{center}\medskip

\begin{center}
{\it Faculty of Physics and Mathematics,\\ Moscow State Regional
University, 105005,\\ Moscow, Radio str., 10A}
\end{center}\medskip

\begin{abstract}
For the one-dimensional linear kinetic
equation  analytical solutions of prob\-lems about
temperature jump and weak evaporation (condensation) over
flat surface are received. The equation has integral of collisions
BGK (Bhatnagar, Gross and Krook) and frequency of collisions of molecules,
affine depending on the module  molecular velocity.


\medskip


\end{abstract}

\begin{center}
\bf  Introduction
\end{center}

In work \cite {1} the linear one-dimensional kinetic
equation with integral of collisions BGK (Bhatnagar, Gross and
Krook) and frequency of colli\-si\-ons, affine depending on the module
velocity of molecules has been entered.

In \cite {1} the theorem about structure of general solution
of the entered equation has been proved.

In the present work which is continuation \cite{1},
exact solutions of the problem about temperature jump and weak evaporation
(condensation) in the rarefied gas are received. These two problems following
\cite{2} we will name the generalized Smoluchowsky' problem , or simply
the Smoluchow\-sky problem.

Let us stop on history of exclusively analytical solutions
of the genera\-li\-zed Smoluchowsky' problem.

For simple (one-nuclear) rarefied gas with a constant
frequency of collisions of molecules the analytical solution
 of the generalized of Smo\-lu\-chow\-sky' problems it is received in \cite{3}.

In \cite {4a} the generalized of Smolu\-chow\-sky' problem  was analytically
solved for simple rarefied gas with frequency of collisions
the molecules,  linearly depending on the module of molecular velocity.
In \cite {4} the problem about strong evaporation (condensation)
with constant frequency of collisions has been analytically solved.

Let us notice, that for the first time the problem
about temperature jump with
frequency of collisions of molecules, linearly
depending on the module
molecular velocity, was analytically solved
by Cassel and Williams in work \cite{5} in 1972.

Then in works \cite{6,7,8} the generalized Smoluchowsky' problem
also analytical solution for case of multinuclear (molecular)
gases has been received.

In works \cite{9,10,11} the problem  about behaviour of the quantum
Boze-gas at low temperatures (similar to the temperature jump problem
for electrons in metal) is considered. We used
the kinetic equation with excitation fonons agrees to N.N. Bogolyubov.

In works \cite{12,13} the problem about temperature jump
for electrons of degenerate plasmas  in metal has been solved.

In work \cite{14} the analytical solution of the  Smoluchowsky' problem
for quantum gases it has been received.

In work of Cercignani and Frezzotti \cite{15} the Smoluchowsky'  problem
it was considered with use of the one-dimensional
kinetic equations. The full analytical solution of
Smoluchowsky' problem with use of Cercignani---Frezotti equation it has
been received in work \cite{16}.

At the same time there is an unresolved problem about temperature jump
and concentration with use of the BGK--equation with
arbitrary dependence of frequency on velocity, in spite of
on obvious importance of the decision of a problem in similar statement.

In the present work attempt to promote in this direction
is made. Here the case of the affine
dependence of collision frequency on molecular velocity in
models of one-dimensional gas is considered.
Model of one-dimensional gas
gave the good consent with the results devoted to the three-dimensional
gas \cite{16}.

Let us start with statement problem.
Then we will give the solution of the  Smoluchowsky'
problem for the one-dimensional kinetic equation with frequency of collisions,
affine depending  on the module of molecular velocity.

\begin{center}
  {\bf 1. Statement of the problem and the basic equations}
\end{center}

Let us begin with the general statement. Let gas occupies
half-space $x> 0$. The surface temperature $T_s $  and
concentration of sated steam of a surface $n_0 $ are set.
Far from a surface gas moves with some velocity $u $,
being velocity of evaporation (or condensation),
also has the temperature gradient
$$
g_T=\Big(\dfrac{d\ln T}{dx}\Big)_{x=+\infty}.
$$

It is necessary to define jumps of
temperature and concentration depending on velocity and
temperature gradient.

In a problem about weak evaporation
it is required to define tempe\-ra\-ture and concentration jumps
depending on velocity, including a tem\-pe\-ra\-ture gradient
equal to zero, and velocity of evaporation (condensation) is enough small.
The last means, that
$$
u \ll v_T.
$$

Here $v_T$ is the heat velocity of molecules, having order
of sound velocity order,
$$
v_T=\dfrac{1}{\sqrt{\beta_s}}, \qquad \beta_s=\dfrac{m}{2k_BT_s},
$$
$m$ is the mass of molecule, $k_B$ is the Boltzmann constant.

In the problem about tem\-pe\-rature jump it is required to define
tem\-pe\-rature and concentration jumps
depending on a temperature gradient, thus
evaporation (condensation) velocity
it is considered equal to zero, and the temperature gradient is
considered  as small.
It means, that
$$
lg_T\ll 1, \qquad l=\tau v_T,\qquad \tau=\dfrac{1}{\nu_0},
$$
where $l$ is the mean free path of gas molecules,
$\tau$ is the mean relaxation time, i.e. time between two
consecutive collisions of molecules.

Let us unite both problems  (about weak evaporation (condensation) and
temperature jump) in one. We will assume that the gradient
of temperature is small (i.e. relative difference
of temperature on length of mean free path is small) and the
velocity of gas in comparison with sound velocity is small. In
this case the problem supposes linearization and  distribution function
it is possible to search in the form
$$
f(x,v)=f_0(v)(1+h(x,v)),
$$
where
$$
f_0(v)=n_s\Big(\dfrac{m}{2\pi k_BT_s}\Big)^{1/2}
\exp \Big[-\dfrac{mv^2}{2k_BT_s}\Big]
$$
is the absolute Maxwellian.

We take the linear kinetic equation which has been written down rather
functions $h(x,v)$, with integral of collisions of
relaxation type, in \cite{1} integral of collisions BGK named also
(Bhatnagar, Gross and Krook), and having the following form
$$
v\dfrac{\partial h}{\partial x}=\nu(v)\Big[A_0+
A_1\dfrac{v}{v_T}+A_2\Big(\dfrac{v^2}{v_T^2}-\beta\Big)-
h(x,v)\Big].
\eqno{(1.1)}
$$

Here
$A_\alpha$ ($\alpha=0,1,2$) is the any constants,
subject to definition from
laws of preservation of number of particles (numerical density),
an momentum and energy, $ \nu(v) $ is the collision frequency affine
depending on module molecular velocity,
$$
\nu(v)=\nu_0\Big(1+\sqrt{\pi}a\sqrt{\dfrac{m}{2k_BT_s}}|v|\Big),
$$
$a$ is the  arbitrary positive paramater, $0\leqslant a<+\infty$.

Strictly speaking, in such model is necessary to take frequency
of collisions in the form
$$
\nu(v)=\nu_0\Big(1+\sqrt{\pi}a\sqrt{\dfrac{m}{2k_BT_s}}|v-u_0(x)|\Big),
$$
where $v-u_0(x)$ is the velocity of a molecule in system of coordinates,
concerning which gas is rested in the given point $x $, $v $ is
the velocity of a molecule in laboratory  system of coordinates, $u_0(x)$
is the  mass velocity of gas in a point $x $ in laboratory system
coordinates. We will consider further, that in linear statement
$|u_0(x)|\ll v_T$.

The right part of the equation (1.1) is the linear integral
of collisions, expanded on collision invariants
$$
\hspace{-2.1cm}\psi_0(v)=1,
$$
$$
\psi_1(v)=\sqrt{\dfrac{m}{2k_BT_s}}v,
$$
$$
\psi_2(v)=\dfrac{mv^2}{2k_BT_s}-\beta.
$$

Let us pass in the equation (1.1) to dimensionless velocity
$$
C=\sqrt{\beta}v=\dfrac{v}{v_T}
$$
and dimensionless coordinate
$$
x'=\nu_0 \sqrt{\dfrac{m}{2k_BT_s}}x=\dfrac{x}{l}
$$

The variable $x'$ let us designate again through $x$.

In the dimensionless variables we will rewrite the equation
(1.1) in the form
$$
C\dfrac{\partial h}{\partial x}=(1+\sqrt{\pi}a|C|)\Big[l_0[h]+
2Cl_1[h]+(C^2-\beta)l_2[h]-h(x,C)\Big].
\eqno{(1.2)}
$$

The constant $ \beta $ is finding from an orthogonality condition
of invariants $ \psi_0(v) $ and $ \psi_2(v) $. Orthogonality here
is understood as equality to zero of scalar product with weight
$\rho(C)=(1+\sqrt{\pi}a|C|)\exp{(-C^2)}$
$$
(f,g)=\int\limits_{-\infty}^{\infty}\rho(C)f(C)g(C)dC.
$$

From here we receive that
$$
\beta=\beta(a)=\dfrac{2a+1}{2(a+1)}.
$$

\begin{center}
\bf 2. Laws of preservation and transformation of the kinetic equation
\end{center}

The modelling integral of collisions should satisfy to laws
preservations of number of particles (numerical density), momentum and
energy
$$
(\psi_\alpha,M[h])\equiv \nu_0\int\limits_{-\infty}^{\infty}
e^{-C^2}(1+\sqrt{\pi}a|C|)M[h]\psi_\alpha(C)dC=0,
\eqno{(2.1)}
$$
where
$$
M[h]=A_0+A_1+(C^2-\beta)A_2-h(x,C).
$$

From the first equation from (2.1), i.e. preservation law of
number of particles
$(\psi_0,M[h])=0$ we receive that
$$
A_0=\dfrac{(1,h)}{(1,1)}=\dfrac{1}{\sqrt{\pi}(a+1)}
\int\limits_{-\infty}^{\infty}e^{-C^2}
(1+\sqrt{\pi}a|C|)h(x,C)dC..
$$

From second equation from (2.1), i.e. preservation law of momentum
$(\psi_1,M[h])=0$ we receive that
$$
A_1=\dfrac{(C,h)}{(C,C)}=\dfrac{2}{\sqrt{\pi}(2a+1)}
\int\limits_{-\infty}^{\infty}e^{-C^2}
(1+\sqrt{\pi}a|C|)Ch(x,C)dC.
$$

From third equation from (2.1), i.e. preservation law of
energy\\
$(\psi_2,M[h])=0$ we receive that
$$
A_2=\dfrac{(C^2-\beta,h)}{(C^2-\beta,C^2-\beta)}=
$$
$$
=\dfrac{4(a+1)}{\sqrt{\pi}(4a^2+7a+2)}
\int\limits_{-\infty}^{\infty}e^{-C^2}
(1+\sqrt{\pi}a|C|)(C^2-\beta)h(x,C)dC.
$$

Let us return to the equation (1.2) and by means of received above equalities
let us transform this equation to the form
$$
C\dfrac{\partial h}{\partial x}+(1+\sqrt{\pi}a|C|)h(x,C)=
$$
$$
=(1+\sqrt{\pi}a|C|)\dfrac{1}{\sqrt{\pi}}\int\limits_{-\infty}^{\infty}
e^{-C'^2}(1+\sqrt{\pi}a|C'|)q(C,C',a)h(x,C')dC'.
\eqno{(2.2)}
$$

Here $q(C,C',a)$ is the kernel of equation, \medskip
$$
q(C,C',a)=r_0(a)+r_1(a)CC'+r_2(a)(C^2-\beta(a))(C'^2-\beta(a)),
$$ \medskip
$$
r_0(a)=\dfrac{1}{a+1},\qquad r_1(a)=\dfrac{2}{2a+1},\qquad
r_2(a)=\dfrac{4(a+1)}{4a^2+7a+2}.
$$ \medskip

\begin{center}
  \bf 3.  Derivation of boundary conditions and the formulation
of boundary problem
\end{center}

Rectilinear substitution it is possible to check up, that the kinetic
equation (2.2) has following four private solutions
$$
h_0(x,C)=1,
$$
$$
h_1(x,C)=C,
$$
$$
h_2(x,C)=C^2,
$$
$$
h_3(x,C)=\Big(C^2-\dfrac{3}{2}\Big)\Big(x-\dfrac{C}{1+\sqrt{\pi}a|C|}\Big).
$$

Let us consider, that molecules are reflected from a wall purely
dif\-fu\-si\-vely, i.e. they are reflected from a wall with Maxwell distribution
by velocities, i.e.
$$
f(x,v)=f_0(v),\qquad v_x>0.
$$
From here we receive for function $h(x,C)$ condition
$$
h(0,C)=0, \qquad C>0.
\eqno{(3.1)}
$$

Condition (3.1) is the first boundary condition to the equation (2.2).

For asymptotic distribution of Chepmen---Enskog we will search in
the form of a linear combination of its partial solutions with
unknown coeffitients
$$
h_{as}(x,C)=A_0+A_1C+A_2\Big(C^2-\dfrac{1}{2}\Big)+$$$$+
A_3\Big(C^2-\dfrac{3}{2}\Big)\Big(x-\dfrac{C}{1+\sqrt{\pi}a|C|}\Big).
\eqno{(3.2)}
$$

We consider the distribution of number density
$$
n(x)=\int\limits_{-\infty}^{\infty}f(x,v)dv=
\int\limits_{-\infty}^{\infty}f_0(v)(1+h(x,v))dv=
n_0+\delta n(x).
$$

Here
$$
n_0=\int\limits_{-\infty}^{\infty}f_0(v)dv,\qquad
\delta n(x)=\int\limits_{-\infty}^{\infty}f_0(v)h(x,v)dv.
$$
From here we receive that
$$
\dfrac{\delta n(x)}{n_0}=\dfrac{1}{\sqrt{\pi}}
\int\limits_{-\infty}^{\infty}e^{-C^2}h(x,C)dC.
$$

We denote
$$
n_e=n_0\dfrac{1}{\sqrt{\pi}}
\int\limits_{-\infty}^{\infty}e^{-C^2}(1+h_{as}(x=0,C))dC.
$$

From here we receive that
$$
\varepsilon_n\equiv \dfrac{n_e-n_0}{n_0}=\dfrac{1}{\sqrt{\pi}}
\int\limits_{-\infty}^{\infty}e^{-C^2}h_{as}(x=0,C)dC.
\eqno{(3.3)}
$$

The quantity $\varepsilon_n$ is the unknown  jump of concentration.

Substituting (3.2) in (3.3), we find, that
$$
\varepsilon_n=A_0.
\eqno{(3.4)}
$$

From definition of dimensional velocity of gas
$$
u(x)=\dfrac{1}{n(x)}\int\limits_{-\infty}^{\infty}f(x,v)vdv
$$
we receive, that in linear approximation dimensional mass velocity
is equal
$$
U(x)=\dfrac{1}{\sqrt{\pi}}\int\limits_{-\infty}^{\infty}
e^{-C^2}h(x,C)CdC.
$$
Setting "far from a wall"\, velocity of evaporation (condensation),
let us write
$$
U=\dfrac{1}{\sqrt{\pi}}\int\limits_{-\infty}^{\infty}
e^{-C^2}h_{as}(x,C)CdC.
\eqno{(3.5)}
$$

Substituting in (3.5) distribution (3.2), we receive, that
$$
A_1=2U+A_3\omega(a),
\eqno{(3.6)}
$$
where $\omega(a)$ is the number parameter,
$$
\omega(a)=\dfrac{2}{\sqrt{\pi}}\int\limits_{-\infty}^{\infty}e^{-C^2}
\dfrac{C^2(C^2-3/2)dC}{1+\sqrt{\pi}a|C|}.
$$

\begin{figure}[ht]
\begin{flushleft}
\includegraphics[width=16.0cm, height=10cm]{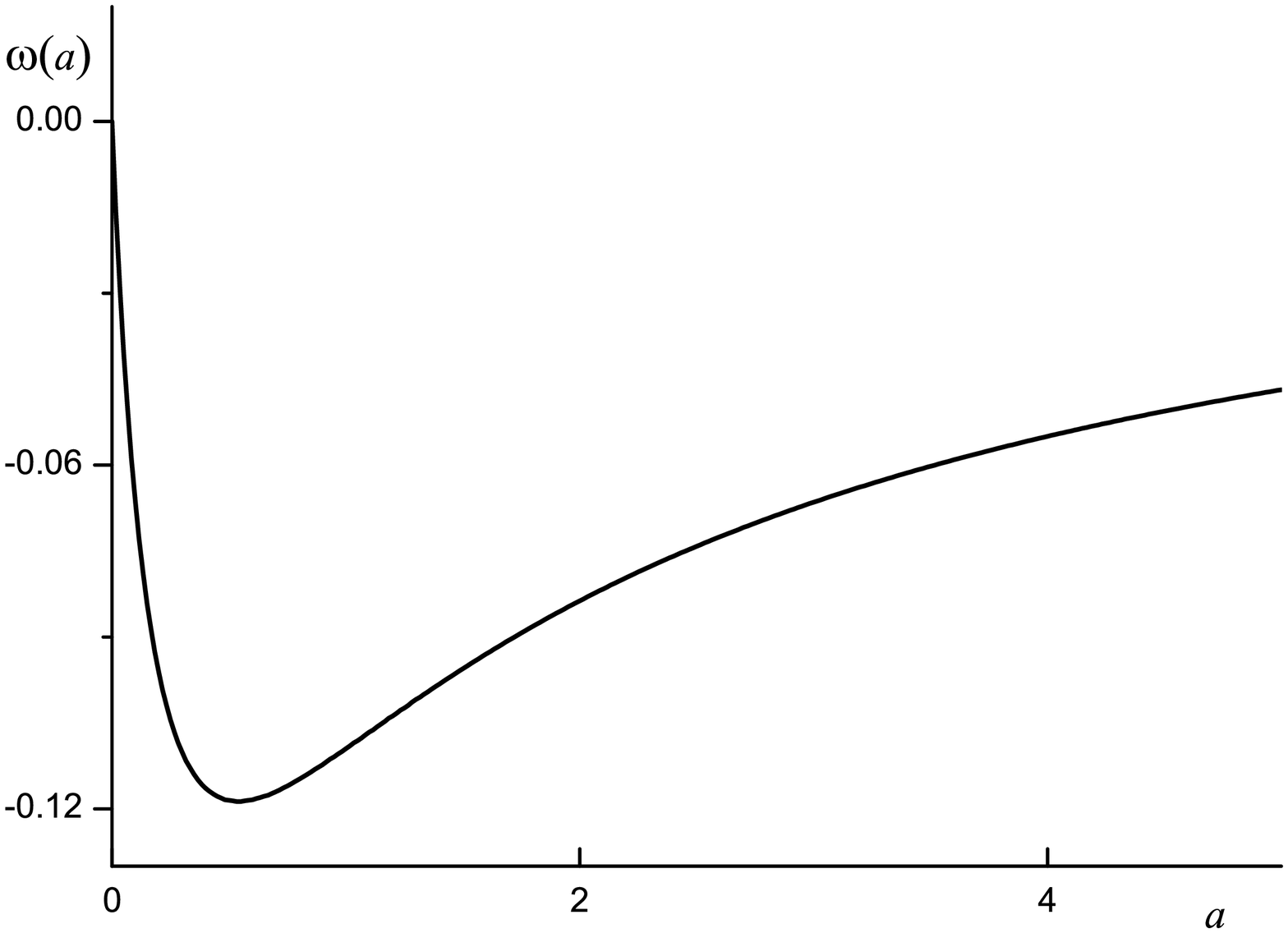}
\end{flushleft}
\center{Fig. 1. Dependence of quantity $ \omega =\omega(a)$ on parameter
of problem $a$.}
\end{figure}

Let us consider temperature distribution
$$
T(x)=\dfrac{2}{kn(x)}\int\limits_{-\infty}^{\infty}\dfrac{m}{2}
(v-u_0(x))^2f(x,v)dv.
$$

From here we find, that
$$
\dfrac{\delta T(x)}{T_0}=-\dfrac{\delta n}{n_0}+\dfrac{2}{\sqrt{\pi}}
\int\limits_{0}^{\infty}e^{-C^2}h(x,C)C^2dC=
$$
$$
=\dfrac{2}{\sqrt{\pi}}
\int\limits_{0}^{\infty}e^{-C^2}h(x,C)(C^2-\dfrac{1}{2})dC.
$$

From here follows, that at $x\to + \infty $ asymptotic
distribution is equal
$$
\dfrac{\delta T_{as}(x)}{T_0}=\dfrac{2}{\sqrt{\pi}}
\int\limits_{0}^{\infty}e^{-C^2}h_{as}(x,C)(C^2-\dfrac{1}{2})dC.
\eqno{(3.7)}
$$

Setting of the gradient of temperature far from a wall means, that
distribution of temperature looks like
$$
T(x)=T_e+\Big(\dfrac{dT}{dx}\Big)_{x=+\infty}\cdot x=T_e+G_Tx,
$$
where
$$
G_T=\Big(\dfrac{dT}{dx}\Big)_{+\infty}.
$$

This distribution we will present in the form
$$
T(x)=T_s\Big(\dfrac{T_e}{T_s}+g_Tx\Big)=T_s\Big(1+
\dfrac{T_e-T_s}{T_s}+g_Tx\Big), \quad x\to +\infty,
$$
where
$$
g_T=\Big(\dfrac{d\ln T}{dx}\Big)_{x=+\infty},
$$
or
$$
T(x)=T_s(1+\varepsilon_T+g_Tx),\qquad x\to +\infty,
$$
where
$$
\varepsilon_T=\dfrac{T_e-T_s}{T_s}
$$
is the unknown temperature jump.

From expression (3.7) is visible, that relative change
of temperature far from a wall is described by linear function
$$
\dfrac{\delta T_{as}(x)}{T_s}=\dfrac{T(x)-T_s}{T_s}=\varepsilon_T+g_Tx,\quad
x\to+\infty
\eqno{(3.8)}
$$

Substituting (3.2) in (3.7), we receive, that
$$
\dfrac{\delta T_{as}(x)}{T_s}=A_2+A_3x.
\eqno{(3.10)}
$$

Comparing (3.7) and (3.10), we find
$$
A_2=\varepsilon_T, \qquad A_3=g_T.
$$

So, asymptotic function of  Chepmen---Enskog' distribution
is const\-ruc\-ted
$$
h_{as}(x,C)=\varepsilon_n+ \varepsilon_T+(2U+\omega(a) g_T)C+
$$
$$
+\Big(C^2-\dfrac{3}{2}\Big)\Big[\varepsilon_T+g_T\Big(x-
\dfrac{C}{1+\sqrt{\pi}a|C|}\Big)\Big].
$$

Now we will formulate the second boundary condition to the equation (2.2)
$$
h(x,C)=h_{as}(x,C)+o(1), \qquad x\to +\infty.
\eqno{(3.11)}
$$

Now we will formulate the basic boundary problem, which is gene\-ralized
Smoluchowsky' problem. This problem consists in finding of  such
solution of the kinetic equation (2.2) which satisfies
to boundary conditions (3.1) and (3.11).

\begin{center}
{\bf 4. Transformation of boundary problem}
\end{center}

In the equation (2.2) we will carry out variable replacement
$ \sqrt {\pi} a\to a $ also we will write down the received
equation in the form
$$
\dfrac{C}{1+a|C|}\dfrac{\partial h}{\partial x}+h(x,C)=$$$$=
\int\limits_{-\infty}^{\infty}e^{-C'^2}(1+a|C'|)q(C,C',a)h(x,C')dC'.
\eqno{(4.1)}
$$

In this equation $q(C,C',a)$ is the kernel of equation,
$$
q(C,C',a)=r_0(a)+r_1(a)CC'+r_2(a)(C^2-\beta(a))(C'^2-\beta(a)),
$$
where
$$
r_0(a)=\dfrac{1}{a+\sqrt{\pi}},\quad r_1(a)=\dfrac{2}{2a+\sqrt{\pi}},
\qquad r_2(a)=\dfrac{4(a+\sqrt{\pi})}{4a^2+7\sqrt{\pi}a+2\pi},
$$
$$
\beta(a)=\dfrac{1}{2}\dfrac{2a+\sqrt{\pi}}{a+\sqrt{\pi}}.
$$

Let us make in the equation (4.1) replacement of a variable
$C=C (\mu),C'=C(\mu')$, where
$$
C(\mu)=\dfrac{\mu}{1-a|\mu|},\qquad |\mu|<\alpha,\qquad \alpha=\dfrac{1}{a}.
$$

Let us designate function $h(x, C(\mu)) $ again through $h(x, \mu) $.
The equation (4.1) passes in the following equation, standard for the
transport theory
$$
\mu\dfrac{\partial h}{\partial x}+h(x,\mu)=\int\limits_{-\alpha}^{\alpha}
\rho(\mu')q(\mu,\mu')h(x,\mu')d\mu',
\eqno{(4.2)}
$$
where
$$
\rho(\mu')=\exp\Big[-\Big(\dfrac{\mu'}{1-a|\mu'|}\Big)^2\Big]
\dfrac{1}{(1-a|\mu'|)^3},
$$
$$
q(\mu,\mu')=r_0(a)+r_1(a)\dfrac{\mu}{1-a|\mu|}\dfrac{\mu'}{1-a|\mu'|}+$$$$+
r_2(a)\Big[\Big(\dfrac{\mu}{1-a|\mu|}\Big)^2-\beta(a)\Big]
\Big[\Big(\dfrac{\mu'}{1-a|\mu'|}\Big)^2-\beta(a)\Big].
$$

Let us notice, that on the ends of the intervals of integration
we have
$$
\rho (\pm \alpha) =0,
$$
and, besides,
$$
\lim\limits _ {\mu\to\pm \alpha} \rho (\mu) C^n (\mu) =0
$$
for any natural $n $.

Boundary conditions (3.1) and (3.11) after variable replacement
$C=C (\mu) $ pass in the following
$$
h(0,\mu)=0, \qquad 0<\mu<\alpha,
\eqno{(4.3)}
$$
and
$$
h(x,\mu)=h_{as}(x,\mu)+o(1),\qquad x\to +\infty,
\eqno{(4.4)}
$$
where
$$
h_{as}(x,\mu)=\varepsilon_n+\varepsilon_T+(2U+ g_T\omega(a))C(\mu)+
$$
$$
+ \Big(C^2(\mu)-\dfrac{3}{2}\Big)\Big[\varepsilon_T+g_T(x-\mu)\Big].
\eqno{(4.5)}
$$

In (4.5) the designation is entered
$$
\omega(a)=\dfrac{2}{\sqrt{\pi}}\int\limits_{-\alpha}^{\alpha}
e^{-C^2(\mu)}C^2(\mu)\Big[C^2(\mu)-\dfrac{3}{2}\Big]\dfrac{d\mu}{1-a|\mu|}.
$$

Let us solve further a boundary problem (4.2) -- (4.4).

\begin{center}
  \bf 5. Eigenfunction and eigenvalues
\end{center}

Seperation of variables in the equation (4.2), taken in the form
$$
h_\eta(x,\mu)=\exp\Big(-\dfrac{x}{\eta}\Big)\Phi(\eta,\mu), \qquad
\eta \in \mathbb{C},
\eqno{(5.1)}
$$
transforms equation (5.1) to characteristic equation
$$
(\eta-\mu)\Phi(\eta,\mu)=\eta Q(\eta,\mu),\qquad \eta,\mu\in (-\alpha,+\alpha),
\eqno{(5.2)}
$$
where
$$
Q(\eta,\mu)=r_0(a)n_0(\eta)+r_1(a)C(\mu)n_1(\eta)+\hspace{4cm}$$$$+
r_2(a)\Big(C^2(\eta)-\beta(a)\Big)\Big(C^2(\mu)-\beta(a)\Big).
$$

Here
$$
n_\alpha(\eta)=\int\limits_{-\alpha}^{\alpha}\Phi(\eta,\mu)C^\alpha(\mu)
\rho(\mu)d\mu,\qquad \alpha=0,1,2,
\eqno{(5.3)}
$$
is the zero, first and second moments of eigen function with
weight $\rho(\mu)$.

Eigen functions of the continuous spectrum filling
by the continuous fashion an interval $ (-\alpha, \alpha) $,
we find \cite{17} in space of the generalized functions
$$
\Phi(\eta,\mu)=\eta Q(\eta,\mu)P\dfrac{1}{\eta-\mu}+g(\eta)\delta(\eta-\mu),
\quad \eta\in (-\alpha,\alpha).
\eqno{(5.4)}
$$

Here $g(\eta)$ is the unknown function, defined from
equations (5.3), $Px^{-1}$  is the distribution, meaning
principal value of integral
by integ\-ra\-tion $x^{-1}$, $\delta(x)$ is the Dirac delta-function.

Let us substitute eigen functions (5.4) in normalization equalities
(5.3). We will receive the following system of the dispersion equations
$$
n_\alpha(\eta)+\eta\int\limits_{-\alpha}^{\alpha}Q(\eta,\mu)C^\alpha(\mu)
\rho(\mu)\dfrac{d\mu}{\mu-\eta}=g(\eta)\rho(\eta)C^\alpha(\eta),
\eqno{(5.5)}
$$
$\alpha=0,1,2$.

We denote
$$
t_n(\eta)=\eta \int\limits_{-\alpha}^{\alpha}C^n(\mu)\dfrac{\rho(\mu)d\mu}
{\mu-\eta},\qquad n=0,1,2,3,4.
$$

Now system of the dispersion equations (5.5) it is possible
transform to the form
$$
n_\alpha(\eta)+r_0(a)n_0(\eta)t_\alpha(a)+r_1(a)n_1(\eta)t_1(\eta)+
$$
$$
+(n_2(\eta)-\beta(a)n_0(\eta))(t_{\alpha+2}(\eta)-\beta(a)t_\alpha(\eta))=
g(\eta)\rho(\eta)C^\alpha(\eta),
\eqno{(5.6)}
$$
where $\alpha=0,1,2$.

Let us write down the equations (5.6) in the vector form
$$
\Lambda(\eta)n(\eta)=g(\eta)\rho(\eta)\left[\begin{array}{c}
                                        1 \\
                                        C(\eta) \\
                                        C^2(\eta)
                                      \end{array}\right].
\eqno{(5.7)}
$$

Here $\Lambda(\eta)$ is the dspersion matrix-function with
elements
$$
\lambda_{ij}(\eta)\quad (i,j=1,2,3),
$$
$n(\eta)$ is the normalization vector with elements $n_\alpha(\eta)\quad
(\alpha=0,1,2)$.

Elements of the dispersion matrix in the explicit form will
more low be necessary
$$
\lambda_{11}(z)=1+\Big[r_2(a)+\beta^2(a)r_2(a)\Big]t_0(z)-
\beta(a)r_2(a)t_2(z),
$$
$$
\lambda_{12}(z)=r_1(a)t_1(z),
$$
$$
\lambda_{13}(z)=r_2(a)\Big[-\beta(a)t_0(z)+t_2(z)\Big],
$$

$$
\lambda_{21}(z)=\Big[r_0(a)+\beta^2(a)r_2(a)\Big]t_1(z)-\beta(a)
r_2(a)t_3(z),
$$
$$
\lambda_{22}(z)=1+r_1(a)t_3(z),
$$
$$
\lambda_{23}(z)=r_2(a)\Big[-\beta(a)t_1(z)+t_3(z)\Big],
$$

$$
\lambda_{31}(z)=\Big[r_0(a)+\beta^2(a)r_2(a)\Big]t_2(z)-\beta(a)
r_2(a)t_4(z),
$$
$$
\lambda_{32}(z)=r_1(a)t_3(z),
$$
$$
\lambda_{33}(z)=1+r_2(a)\Big[-\beta(a)t_2(z)+t_4(z)\Big].
$$

We introduce the dispersion function $\lambda(z)$, $\lambda(z)=\det
\Lambda(z)$. In the explicit form we have
$$
\lambda(z)=\lambda_{11}(z)\lambda_{22}(z)\lambda_{33}(z)+
r_1(a)t_3(z)\lambda_{13}(z)\lambda_{21}(z)+
$$

$$
+r_1(a)t_1(z)\lambda_{31}(z)\lambda_{23}(z)-
\lambda_{13}(z)\lambda_{22}(z)\lambda_{31}(z)-
$$

$$
-r_1(a)t_3(z)\lambda_{11}(z)\lambda_{23}(z)-r_1(a)t_1(z)\lambda_{21}(z)
\lambda_{33}(z).
$$

From vector equation (5.7) we find
$$
n_\alpha(\eta)=g(\eta)\rho(\eta)\dfrac{\Lambda_\alpha(\eta)}{\lambda(\eta)},
\quad \alpha=0,1,2,
\eqno{(5.8)}
$$
where $\Lambda_\alpha(\eta)$  is the determinant received from
determinant of system (5.6) by replacement in it $ \alpha $-th
column by the column from free members of this system.
We will write out these
determinants in the explicit form
$$
\Lambda_0(z)=\Lambda_{11}(z)-C(z)\Lambda_{21}(z)+C^2(z)\Lambda_{31}(z)=
\lambda_{22}(z)\lambda_{33}(z)-
$$
$$
-r_1(a)t_3(z)\lambda_{23}(z)-C(z)r_1(a)\Big[t_1(z)\lambda_{33}(z)-
t_2(z)\lambda_{13}(z)\Big]+
$$
$$
+C^2(z)\Big[r_1(a)t_1(z)\lambda_{23}(z)-
\lambda_{22}(z)\lambda_{13}(z)\Big],
$$

$$
\Lambda_1(z)=\Lambda_{12}(z)+C(z)\Lambda_{22}(z)-C^2(z)\Lambda_{32}(z)=
-\lambda_{21}(z)\lambda_{33}(z)+
$$
$$
+\lambda_{31}(z)\lambda_{33}(z)+C(z)\Big[\lambda_{11}(z)\lambda_{33}(z)-
\lambda_{31}(z)\lambda_{13}(z)\Big]-
$$
$$
-C^2(z)\Big[\lambda_{11}(z)\lambda_{23}(z)-
\lambda_{21}(z)\lambda_{13}(z)\Big],
$$

$$
\Lambda_2(z)=\Lambda_{31}(z)-C(z)\Lambda_{32}(z)+C^2(z)\Lambda_{33}(z)=
$$
$$
=r_1(a)t_3(z)\lambda_{21}(z)-\lambda_{31}(z)\lambda_{22}(z)-
C(z)r_1(a)\Big[t_3(z)\lambda_{11}(z)-t_1(z)\lambda_{33}(z)\Big]+
$$
$$
+C^2(z)\Big[\lambda_{11}(z)\lambda_{22}(z)-r_1(a)
t_1(z)\lambda_{21}(z)\Big].
$$

Here $\Lambda_{ij}(z)$ is the minor of element $\lambda_{ij}(z)$.

By means of equalities (5.8) we will transform equality for $Q(\eta, \mu) $
to the form
$$
Q(\eta,\mu)=\tilde Q(\eta,\mu)\dfrac{g(\eta)}{\lambda(\eta)}\rho(\eta),
\eqno{(5.9)}
$$
where
$$
\tilde Q(\eta,\mu) =r_0(a)\Lambda_0(\eta)+r_1(a)C(\mu)\Lambda_1(\eta)+
\hspace{4cm}
$$
$$
+r_2(a)\Big[C^2(\mu)-\beta(a)\Big]\Big[\Lambda_2(\eta)-\beta(a)
\Lambda_0(\eta)\Big].
$$

By means of equality (5.9) we will transform expression (5.4) for
eigen functions
$$
\Phi(\eta,\mu)=\tilde \Phi(\eta,\mu)g(\eta),
\eqno{(5.10)}
$$
where
$$
\tilde \Phi(\eta,\mu)=\eta\dfrac{\tilde Q(\eta,\mu)}{\lambda(\eta)}\rho(\eta)
P\dfrac{1}{\eta-\mu}+\delta(\eta-\mu).
\eqno{(5.11)}
$$

From equality (5.10) it is visible, that eigen functions are defined
accurate within to coefficient -- any function $g (\eta) $,
identically not equal to zero. Owing to uniformity of the initial
kinetic equation it is possible to consider this function identically
equal to unit ($g (\eta) \equiv 1$) and further in quality
eigen function corresponding to  continuous spectrum, it is possible
to consider the functions defined by equality (5.11).
Apparently from the solution of the characteristic equation, continuous
spectrum of the characteristic equation is the set
$$
\sigma_c=\{\eta: -\alpha<\eta<+\alpha\}.
$$

By definition the discrete spectrum of the characteristic
equation consists of set of zero of dispersion function.

Expanding dispersion function in Laurent series in a vicinity
infinitely remote point, we are convinced, that it in this point
has zero of the fourth order. Applying an argument principle \cite{18}
from the theory of functions complex variable, it is possible to show, that
other zero, except $z_i =\infty $, dispersion function not
has. Thus, the discrete spectrum of the characteristic
equations consists of one point $z_i =\infty $,
multiplication factor which it is equal four,
$$
\sigma_d=\{z_i=\infty\}.
$$

To point $z_i =\infty $, as to the 4-fold point of  discrete spectrum,
corresponds following four discrete (partial) solutions
of the kinetic decision (4.2)
$$
h_0(x,\mu)=1,
$$
$$
h_1(x,\mu)=C(\mu),
$$
$$
h_2(x,\mu)=C^2(\mu)-\dfrac{1}{2},
$$
$$
h_3(x,\mu)=(x-\mu)\Big(C^2(\mu)-\dfrac{3}{2}\Big).
$$

Let us result formulas Sokhotsky for the difference and the sum of the boundary
values of dispersion function from above and from below on the
$(-\alpha,+\alpha)$
$$
\lambda^+(\mu)-\lambda^-(\mu)=2\pi i \rho(\mu) \tilde Q(\mu,\mu),\quad
\mu\in (-\alpha,+\alpha),
$$
and
$$
\dfrac{\lambda^+(\mu)+\lambda^-(\mu)}{2}=\lambda(\mu),\quad
\mu\in (-\alpha,+\alpha).
$$

\begin{center}
\bf 6. Analytical solution of boundary value problem
\end{center}

Here we will prove the theorem about analytical solution of the basic
boundary problem (4.2) - (4.4).

{\sc Theorem.} {\it The boundary problem (4.2---(4.4) has
the unique solution, representable in the form of the sum linear
combinations of discrete (partial) solutions of this equation
and integral on the continuous spectrum
on eigenfunctions correspond to the continuous spectrum
$$
h(x,\mu)=h_{as}(x,\mu)+\int\limits_{0}^{\alpha}
\exp\Big(-\dfrac{x}{\eta}\Big)F(\eta,\mu)A(\eta)d\eta.
\eqno{(6.1)}
$$

In equality (6.1) $ \varepsilon_n $ and $ \varepsilon_T $ are
unknown coefficients (of discrete spectrum), $U $ and $g_T $
are the given values, $A (\eta) $ is the unknown function
(coefficient of the continuous spectrum).}

Coefficients of discrete and continuous spectra are subject
to finding from boundary conditions.

Expansion (6.1) it is possible to present in classical sense
$$
h(x,\mu)=\varepsilon_n+\varepsilon_T+(2U+ g_T\omega)C(\mu)+
\Big(C^2(\mu)-\dfrac{3}{2}\Big)\Big[\varepsilon_T+
$$
$$
+g_T(x-\mu)\Big]+e^{-x/\mu}A(\mu)+\int\limits_{0}^{\alpha}
e^{-x/\eta}\dfrac{\eta R(\eta,\mu)\rho(\eta)A(\eta)}
{\lambda(\eta)(\eta-\mu)}d\eta.
\eqno{(6.1')}
$$

{\sc Proof.}
Let us substitute decomposition (6.1) in a boundary condition (4.2).
We receive the integral equation
$$
h_{as}(0,\mu)+\int\limits_{0}^{\alpha}F(\eta,\mu)A(\eta)d\eta=0,\quad
0<\mu<\alpha.
$$

In an explicit form this equation looks like
$$
h_{as}(0,\mu)+\int\limits_{0}^{\alpha}\dfrac{\eta R(\eta,\mu)\rho(\eta)}
{\lambda(\eta)(\eta-\mu)}A(\eta)d\eta+A(\mu)=0,\quad 0<\mu<\alpha.
\eqno{(6.2)}
$$

Let us enter auxiliary function
$$
N(z)=\int\limits_{0}^{\alpha}\dfrac{\eta R(\eta,z)\rho(\eta)}
{\lambda(\eta)(\eta-z)}A(\eta)d\eta,
\eqno{(6.3)}
$$
for which according to formulas Sohotsky we have
$$
N^+(\mu)-N^-(\mu)=2\pi i \mu\dfrac{R(\mu,\mu)}{\lambda(\mu)}
\rho(\mu)A(\mu),\quad 0<\mu<\alpha,
\eqno{(6.4)}
$$
$$
\dfrac{N^+(\mu)+N^-(\mu)}{2}=\int\limits_{0}^{\alpha}
\dfrac{\eta R(\eta,\mu)\rho(\eta)}
{\lambda(\eta)(\eta-\mu)}A(\eta)d\eta, \quad 0<\mu<\alpha.
\eqno{(6.5)}
$$

Let us transform the equation (6.2) according to equalities (6.4) and (6.5)
$$
h_{as}(0,\mu)+\dfrac{N^+(\mu)+N^-(\mu)}{2}+\lambda(\mu)
\dfrac{N^+(\mu)-N^-(\mu)}{2\pi i \rho(\mu) R(\mu,\mu)}=0,
$$
whence
$$
2 \pi i \mu \rho(\mu)R(\mu,\mu)h_{as}(0,\mu)+
\pi i \mu \rho(\mu)R(\mu,\mu)[N^+(\mu)+N^-(\mu)]+
$$
$$
+\lambda(\mu)[N^+(\mu)-N^-(\mu)]=0,\qquad 0<\mu<\alpha.
\eqno{(6.6)}
$$

Considering formulas of Sokhotsky for dispersion function,
let us transform the equation (6.6) to the non-uniform Riemann'
boundary value problem
$$
\lambda^+(\mu)[N^+(\mu)+h_{as}(0,\mu)]-$$$$-
\lambda^-(\mu)[N^-(\mu)+h_{as}(0,\mu)]=0,\quad 0<\mu<\alpha.
\eqno{(6.7)}
$$

Let us consider the corresponding homogeneous boundary value problem
$$
\dfrac{X^+(\mu)}{X^-(\mu)}=\dfrac{\lambda^+(\mu)}{\lambda^-(\mu)},
\quad 0<\mu<\alpha.
\eqno{(6.8)}
$$

The solution of this problem which is limited and not disappearing
in points $z=0$ and $z =\alpha $ we will take without a derivation
$$
X(z)=\dfrac{1}{z^2}\exp V(z),
\eqno{(6.9)}
$$
where
$$
V(z)=\dfrac{1}{\pi}\int\limits_{0}^{\alpha}\dfrac{\theta(\mu)-2\pi}
{\mu-z}d\mu,
\eqno{(6.10)}
$$
where $ \theta(\mu)=\arg\lambda^+(\mu) $ is the principal value
of the argument, fixed in zero by condition $ \theta(0) =0$.

Let us transform the problem (6.7) by means of homogeneous problem (6.8) to
the problem of  finding of analytical function on its jump on the cut,
$$
X^+(\mu)[N^+(\mu)+h_{as}(0,\mu)]=
$$
$$
=X^-(\mu)[N^-(\mu)+h_{as}(0,\mu)],\qquad 0<\mu<\alpha.
\eqno{(6.11)}
$$

Let us find singularities of boundary  condition (6.11).
We will return to function
$N (z) $, which boundary values in the interval $ (0, \alpha) $
enter into the boundary condition (6.7) and which is defined by equality
(6.3). We will present this function in the explicit form
$$
N(z)=r_0(a)P(z)+r_1(a)\dfrac{z}{1-az}Q(z)+
r_2(a)\Big[\Big(\dfrac{z}{1-az}\Big)^2-\beta(a)\Big]R(z),
\eqno{(6.12)}
$$
where
$$
P_n(z)=\int\limits_{0}^{\alpha}
\eta\dfrac{\Lambda_n(\eta)}{\lambda(\eta)}\rho(\eta)
\dfrac{A(\eta)d\eta}{\eta-z},\qquad n=0,1,2,
$$
$$
Q(z)=\int\limits_{0}^{\alpha}
\eta\dfrac{\Lambda_1(\eta)}{\lambda(\eta)}\rho(\eta)
\dfrac{A(\eta)d\eta}{\eta-z}\equiv P_1(z),
$$
$$
R(z)=\int\limits_{0}^{\alpha}\eta\dfrac{\Lambda_2(\eta)-
\beta(a)\Lambda_0(\eta)}{\lambda(\eta)}\rho(\eta)
\dfrac{A(\eta)d\eta}{\eta-z} \equiv P_2(z)-\beta(a)P_0(z),
$$

Let us transform equality (6.12), allocating its polar
singularity $\dfrac{z}{1-az}$:
$$
N(z)=r_0(a)P_0(z)-r_2(a)\beta(a)[P_2(z)-\beta(a)P_0(z)]+
$$
$$
+r_1(a)\dfrac{z}{1-az}P_1(z)+r_2(a)\Big(\dfrac{z}{1-az}\Big)^2
[P_2(z)-\beta(a)P_0(z)].
\eqno{(6.13)}
$$

Let us write down equality (6.13) on degrees of its polar singularity
$$
N(z)=P(z)+r_1(a)\dfrac{z}{1-az}Q(z)+r_2(a)\Big(\dfrac{z}{1-az}\Big)^2
R(z),
\eqno{(6.14)}
$$
where
$$
P(z)=[r_0(a)+\beta^2(a)r_2(a)]P_0(z)-r_2(a)\beta(a)P_2(z),
$$
$$
Q(z)\equiv P_1(z),\qquad R(z)=P_2(z)-\beta(a)P_0(z).
$$

From expression (6.14) it is visible, that the boundary
condition (6.11) has double pole in the point $z=1/a $.
Therefore, multiplying this boundary condition on $(1-a\mu)^2$,
we receive
$$
X^+(\mu)[M^+(\mu)+(1-a\mu)^2h_{as}(0,\mu)]=
$$
$$
=X^-(\mu)[M^-(\mu)+(1-a\mu)^2h_{as}(0,\mu)],\quad 0<\mu<\alpha,
\eqno{(6.15)}
$$
where
$$
M(z)=(1-az)^2N(z).
\eqno{(6.16)}
$$

Considering behaviour of the functions entering into the
boundary condition (6.15), we receive the general solution of
the corresponding boundary problem
$$
M(z)=-(1-az)^2h_{as}(0,z)+\dfrac{C_0+C_1z}{X(z)},
\eqno{(6.17)}
$$
where $C_0$ and $C_1$ are arbirary constants, though in this
equality
$$
(1-az)^2h_{as}(0,z)=
$$
$$
=(1-az)^2(\varepsilon_n+\varepsilon_T)-(1-az)z(2U+\omega(a)g_T)-
\Big[z^2-\dfrac{3}{2}(1-az)^2\Big](\varepsilon_n-g_Tz).
$$ \medskip

Let us notice, that the solution (6.17) has in infinitely removed
point $z =\infty $ a pole of the third order, while function
$M(z)$, defined by equality (6.16), has in this point a pole
the first order. That the solution (6.17) could be accepted in
quality of function $M(z)$, defined by equality (6.16), we will lower
order of a pole at the solution (6.17) from three to unit, and then
let us equate coefficients at $z^m (m=1,0)$ in expansion in
point vicinities $z =\infty $ the left and right parts of equality
(6.17). The last is caused by that these coefficients in both
equality parts contain unknown parametres.

Let us expand both the left and right parts of the solution
(6.17) in Laurent series
in a vicinity of infinitely remote point
$$
M_1z+M_0+o(1)=C_1z^3+(C_0-V_1)z^2+(C_1U_2-C_0V_1)z+
$$
$$
+(C_1U_3-C_0U_2)-\varepsilon_n+\dfrac{1}{2}\varepsilon_T+
(2a\varepsilon_n-a\varepsilon_T-2U-g_T\omega(a)-\dfrac{3}{2}g_T)z+
$$
$$
+\Big[-a^2\varepsilon_n+\varepsilon_T(\dfrac{a^2}{2}-1)+2aU+
a(\omega(a)+3))g_T+\Big]z^2+
$$
$$
+\Big(1-\dfrac{3}{2}a^2\Big)g_Tz^3+o(1), \qquad z\to \infty.
\eqno{(6.18)}
$$\medskip

At a derivation of equality (6.18) expansion has been used
$$
V(z)=\dfrac{V_1}{z}+\dfrac{V_2}{z^2}+\cdots, \qquad z\to \infty.
$$

Here
$$
V_n=-\dfrac{1}{\pi}\int\limits_{0}^{\infty}\tau^{n-1}[\theta(\tau)-
2\pi]d\tau, \qquad n=1,2,\cdots.
$$

Besides, at the derivation (6.18) one more has been used
expansion
$$
\exp\Big[-V(z)\Big]=1+\dfrac{V_1^*}{z}+\dfrac{V_2^*}{z^2}+\cdots, \qquad
z\to \infty.
\eqno{(6.19)}
$$

Coefficients $V_n^*$ are expressed through coefficients $V_n $ with
the help of the equality found on the basis of (6.19)
$$
V'(z)=\dfrac{\dfrac{V_1^*}{z^2}+2\dfrac{V_2^*}{z^3}+
3\dfrac{V_3^*}{z^4}\cdots}{1+\dfrac{V_1^*}{z}+\dfrac{V_2^*}{z^2}+
\dfrac{V_3^*}{z^3}\cdots}.
$$

Really, substituting in the left part of this equality the series for
$V(z)$, we receive
$$
-\Big(\dfrac{V_1}{z^2}+2\dfrac{V_2}{z^3}+3\dfrac{V_3}{z^4}\cdots\Big)
\Big(1+\dfrac{V_1^*}{z}+\dfrac{V_2^*}{z^2}+
\dfrac{V_3^*}{z^3}\cdots\Big)=
$$
$$
=\dfrac{V_1^*}{z^2}+2\dfrac{V_2^*}{z^3}+3\dfrac{V_3^*}{z^4}\cdots.
$$

From here we find that
$$
V_1^*=-V_1,\qquad V_2^*=-V_2+\dfrac{1}{2}V_1^2,
$$
$$
V_3^*=-V_3+V_1V_2-\dfrac{1}{6}V_1^3, \cdots.
$$

Equating to zero in the right part of equality (6.18) coefficients
at $z^3$ and $z^2$, we find
$$
C_1=g_T\Big(\dfrac{3}{2}a^2-1\Big),
\eqno{(6.20)}
$$
$$
C_0=g_T\Big[\Big(\dfrac{3}{2}a^2-1\Big)V_1-a(\omega(a)+3)\Big]+$$$$+
a^2\varepsilon_n+\Big(1-\dfrac{a^2}{2}\Big)\varepsilon_T-2aU.
\eqno{(6.21)}
$$

Equating now coefficients in (6.18) at the left and on the right at $z $
and $z^0$, we find
$$
M_1=C_1V_2^*-C_0V_1+2a\varepsilon_n-2U-(\omega(a)+\dfrac{3}{2})g_T,
\eqno{(6.22)}
$$
$$
M_2=C_1V_3^*+C_0V_2^*-\varepsilon_n+\dfrac{1}{2}\varepsilon_T.
\eqno{(6.23)}
$$

Substituting in equalities (6.22) and (6.23) coefficients $C_0$ and $C_1$,
defined accordingly equalities (6.20) and (6.21), we will receive the equ\-ations,
from which unequivocally searched $ \varepsilon_n $ and $ \varepsilon_T $.
Thus, free para\-met\-res $C_0$ and $C_1$ from solution $M(z)$
are found unequivocally,
and also coefficients of the discrete  spectrum are found unequivocally
$ \varepsilon_n $ and $ \varepsilon_T $ of expansion (6.1).
Coefficient of continuous spectrum $A (\eta) $
is searched on the basis of formulas Sokhotsky, applied to
functions $M(z)$, the defined equalities (6.16) and (6.17)
$$
M+(\mu)-M^-(\mu)=2\pi i \mu (1-a\mu)^2\dfrac{R(\mu,\mu)}{\lambda(\mu)}
\rho(\mu)A(\mu)
$$
and
$$
M+(\mu)-M^-(\mu)=(C_0+C_1\mu)
\Big(\dfrac{1}{X^+(\mu)}-\dfrac{1}{X^-(\mu)}\Big).
$$

From these equalities we find
$$
\dfrac{\eta A(\eta)\rho(\eta)}{\lambda(\eta)}=\dfrac{C_0+C_1\eta}
{2\pi i (1-a\eta)^2R(\eta,\eta)}
\Big(\dfrac{1}{X^+(\eta)}-\dfrac{1}{X^-(\eta)}\Big).
\eqno{(7.24)}
$$\medskip

So, all coefficients of expansion (6.1) are established. On
construction, expansion (6.1) satisfies to boundary conditions
(4.3) and (4.4). That fact, that expansion (6.1) satisfies
to the equation (4.2), it is checked directly.

Uniqueness of decomposition (6.1) is proved by a method from
the opposite. The theorem is proved.

\begin{center}
  {\bf 7.  Temperature jump and weak evaporation (condensation)}
\end{center}

Let us return to the decision of the put physical problems.
Coefficient of continuous spectrum (6.24) we will substitute at
first in equality (6.3) for functions $N(z)$, and then we will
take advantage of equality (6.16), which
let us present in the form of the sum
$$
M(z)=C_0K(z)+C_1L(z),
\eqno{(7.1)}
$$
where
$$
K(z)=\dfrac{(1-az)^2}{2\pi i}\int\limits_{0}^{\alpha}
\dfrac{R(\eta,z)}{(1-a\eta)^2R(\eta,\eta)}\Big[\dfrac{1}{X^+(\eta)}-
\dfrac{1}{X^-(\eta)}\Big]\dfrac{d\eta}{\eta-z},
\eqno{(7.2)}
$$
and
$$
L(z)=\dfrac{(1-az)^2}{2\pi i}\int\limits_{0}^{\alpha}
\dfrac{\eta R(\eta,z)}{(1-a\eta)^2R(\eta,\eta)}\Big[\dfrac{1}{X^+(\eta)}-
\dfrac{1}{X^-(\eta)}\Big]\dfrac{d\eta}{\eta-z}.
\eqno{(7.3)}
$$

According to (7.1) we have
$$
M(z)=(C_0K_1+C_1L_1)+(C_0K_0+C_1L_1)+o(1), \qquad z\to \infty,
\eqno{(7.4)}
$$
i.e.
$$
M_1=C_0K_1+C_1L_1,\qquad M_0=C_0K_0+C_1L_0.
\eqno{(7.5)}
$$

The found coefficients $M_1$ and $M_2$ we will substitute in equalities
(6.22) and (6.23), then we will replace in them $C_0$ and $C_1$ it agree
(6.20) and (6.21).
We receive system from two equations rather $ \varepsilon_T $
and $ \varepsilon_n $:
$$
-\Big[a+\Big(1-\dfrac{a^2}{2}\Big)(V_1+K_1)\Big]\varepsilon_T+
[2a-a^2(V_1+K_1)]\varepsilon_n=
$$
$$
=2U[1-a(V_1+K_1)]+g_T\Big[\dfrac{3}{2}+\omega+\Big(\dfrac{3}{2}-1\Big)
(L_1-$$$$-V_2^*+V_1^2+V_1K_1)-(3a-\omega(a))(V_1K_1)\Big]
\eqno{(7.6)}
$$
and
$$
\Big[\dfrac{1}{2}+\Big(1-\dfrac{a^2}{2}\Big)(V_2^*-K_0)\Big]\varepsilon_T+
[-1+a^2(V_2^*-K_0)]\varepsilon_n=
$$
$$
=2aU(V_2^*-K_0)+g_T\Big[\Big(\dfrac{3}{2}a^2-1\Big)(V_1K_0-V_1V_2-
$$
$$
-V_3+L_0)+(3a+\omega(a))(V_2+K_0)\Big].
\eqno{(7.7)}
$$

The basic determinant of this system is equal
$$
\Delta=V_1+K_1-2a(V_2^*-K_0).
$$

Let us bring expressions for other determinants
$$
\Delta_{T,U}=-1+a(V_1+K_1)-a^2(V_2^*-K_0),
$$
$$
\Delta_{n,U}=-\dfrac{1}{2}+\dfrac{1}{2}a(V_1+K_1)-\Big(1+\dfrac{a^2}{2}\Big)
(V_2^*-K_0),
$$
$$
\Delta_{T,g_T}=[1-a^2(V_2^*-K_0)]\Big[\Big(\dfrac{3}{2}a^2-1\Big)
(V_2^*-L_1)-\omega(a)-\dfrac{3}{2}\Big]+
$$
$$
+\Big[\Big(\dfrac{3}{2}a^2-1\Big)V_1-3a-\omega(a)\Big]
[2a(V_2^*-K_0)-V_1-K_1]+
$$
$$
+\Big(\dfrac{3}{2}a^2-1\Big)(V_3^*-L_0)[2a-a^2(V_1+K_1)],
$$\medskip
$$
\Delta_{n,g_T}=-\Big[a+\Big(1-\dfrac{a^2}{2}\Big)(V_1+K_1)\Big]
\Big[(3a+\omega(a))(V_2^*-K_0)-\Big(\dfrac{3}{2}a^2-1\Big)\times
$$
$$
\times(V_3^*-L_0+V_1V_2^+-V_1K_0)\Big]-\Big[\dfrac{1}{2}+
\Big(1-\dfrac{a^2}{2}\Big)(V_2^*-K_0)\Big]\times
$$
$$
\times\Big[\dfrac{3}{2}+\omega(a)+\Big(\dfrac{3}{2}a^2-1\Big)
(L_1-V_2^*+V_1^2+V_1K_1)-(3a+\omega(a))(V_1+K_1)\Big].
$$

We will write down the solution of system  of the equations
(7.6) and (7.7) in the form
$$
\varepsilon_T=2U\dfrac{\Delta_{T,U}}{\Delta}+g_T\dfrac{\Delta_{T,g_T}}
{\Delta},
\eqno{(7.8)}
$$
and
$$
\varepsilon_n=2U\dfrac{\Delta_{n,U}}{\Delta}+g_T\dfrac{\Delta_{n,g_T}}
{\Delta}.
\eqno{(7.9)}
$$

\begin{center}
  \bf 8. Conclusion
\end{center}

In the present work the analytical solution
of boundary problems for the one-dimensional kinetic equation with
frequency of collisions of molecules, affine depending on the module
molecular velocity \cite{1}  is considered. The solution
of the generalized Smoluchowsky' problem about temperature jump and weak
evaporation (condensation)  is considered.

The theorem of expansion of the solution of the  Smoluchowsky'
problem on eigenfunctions of the corresponding characteristic
equation is proved .

\end{document}